\documentclass{Interspeech}

\interspeechcameraready

\title{Modality-Specific Speech Enhancement and Noise-Adaptive Fusion for Acoustic and Body-Conduction Microphone Framework}

\author[affiliation={1}]{Yunsik}{Kim}
\author[affiliation={1,2,3*}]{Yoonyoung}{Chung}
\affiliation{Department of Electrical Engineering, $^{2}$Department of Semiconductor Engineering, $^{3}$Center for Semiconductor Technology Convergence}{Pohang University of Science and Technology (POSTECH)}{Pohang, Korea}

\email{ys.kim@postech.ac.kr, ychung@postech.ac.kr}

\usepackage{comment}
\usepackage{multirow}
\usepackage{makecell}

\begin{document}

\maketitle

\begin{abstract}
    Body-conduction microphone signals (BMS) bypass airborne sound, providing strong noise resistance. However, a complementary modality is required to compensate for the inherent loss of high-frequency information. In this study, we propose a novel multi-modal framework that combines BMS and acoustic microphone signals (AMS) to achieve both noise suppression and high-frequency reconstruction. Unlike conventional multi-modal approaches that simply merge features, our method employs two specialized networks: a mapping-based model to enhance BMS and a masking-based model to denoise AMS. These networks are integrated through a dynamic fusion mechanism that adapts to local noise conditions, ensuring the optimal use of each modality’s strengths. We performed evaluations on the TAPS dataset, augmented with DNS-2023 noise clips, using objective speech quality metrics. The results clearly demonstrate that our approach outperforms single-modal solutions in a wide range of noisy environments.
\end{abstract}

\section{Introduction}
Recent advances in deep learning-based speech enhancement have demonstrated remarkable progress through two dominant methodologies: masking-based and mapping-based approaches \cite{yuliani2021speech}. Masking-based methods \cite{hu2020dccrn, lv2021dccrn+, hao2021fullsubnet, chen2022fullsubnet+, fu2019metricGAN, fu2021metricgan+} primarily operate in the time-frequency (TF) domain, where noisy speech spectrograms—computed via short-time Fourier transform (STFT)—are multiplied by estimated ideal ratio masks to suppress noise components. These methods excel at preserving speech phase information and enabling interpretable TF bin-wise processing. In contrast, mapping-based methods \cite{defossez2020real, rethage2018wavenet, hsieh2020wavecrn, wang2021tstnn, kim2021se} directly estimate clean speech features from noisy inputs. Among them, time-domain end-to-end approaches operate directly on raw waveforms, jointly optimizing both magnitude and phase to reduce spectral decomposition artifacts \cite{defossez2020real, kim2021se}. TF-domain mapping methods, on the other hand, estimate clean spectral magnitudes while reusing the noisy phase, allowing a simpler structure \cite{tan2019learning}.

Despite their strengths, both paradigms face limitations in extreme noise scenarios. Masking-based models suffer from irreversible spectral distortions when noise dominates speech energy \cite{fu2019metricGAN}, while time-domain networks struggle to separate overlapping speech and noise temporal features \cite{kim2021se}. These limitations stem from their reliance on acoustic microphone signals, which inherently capture both speech and noise. Body-conduction microphones, which detect speech-induced vibrations through body tissues, provide inherent noise resistance by avoiding airborne sound transmission \cite{mehta2015using,zheng2003air,quatieri2006exploiting, kang2018transparent}. However, body-conducted speech experiences severe high-frequency attenuation above 2 kHz due to tissue filtering, resulting in muffled and less intelligible speech \cite{shin2012survey, tran2013effect}. Although mapping-based enhancement techniques \cite{defossez2020real, rethage2018wavenet, kim2021se} can partially restore high-frequency content, the irreversible spectral loss reduces their effectiveness. Consequently, existing solutions struggle to perform consistently across the full range of noise conditions, from moderate to extreme noise environments, underscoring the need for a multi-modal approach that dynamically adapts to varying noise environments encountered in real-world scenarios \cite{wang2022multi}.

Recent multi-modal frameworks have demonstrated promising performance by leveraging the complementary strengths of body-conduction microphone signals (BMS) and acoustic microphone signals (AMS). Yu et al.\cite{yu2020time} proposed a convolutional neural network (CNN) that uses BMS as an auxiliary modality to enhance AMS. Wang et al.\cite{wang2022attention} introduced an attention-based fusion mechanism within a convolutional recurrent network, enabling efficient speech enhancement in the complex spectral domain. Wang et al.\cite{Wang2022fusing} reported an attention-based fusion approach with a semi-supervised CycleGAN framework, effectively utilizing limited paired AMS and BMS data. Unlike prior approaches that rely on implicit fusion of BMS and AMS through concatenation or attention-based fusion, we emphasize the need for modality-specific processing to achieve effective spectral restoration and noise suppression.

To achieve this, we introduce the Body-Acoustic Fusion Network (BAF-Net), a multi-modal framework designed to leverage the unique strengths of BMS and AMS through task-specific architectures and a noise-adaptive fusion. Unlike previous approaches, the BAF-Net uses two specialized networks: a mapping-based SE-conformer \cite{kim2021se} to restore high-frequency components in BMS and a masking-based DCCRN \cite{hu2020dccrn} to suppress noise in AMS. Critically, the BAF-Net introduces a mask-guided fusion mechanism that dynamically adjusts modality contributions based on noise conditions inferred from the DCCRN’s output, eliminating the need for explicit signal-to-noise ratio (SNR) estimation. This design allows optimal exploitation of BMS’s inherent noise robustness in high-noise regions and AMS’s spectral richness in low-noise conditions.

The framework’s principal innovations include: (1) task-specialized enhancement, where separate networks explicitly handle BMS’s high-frequency reconstruction and AMS’s noise suppression, and (2) adaptive signal-level fusion, which utilizes fusion coefficients estimated by a lightweight CNN from the noise mask, enabling seamless modality switching. To evaluate our model, we used a simulated dataset built from the TAPS corpus \cite{kim2025tapsthroatacousticpaired} and DNS-2023 noise data \cite{dubey2023icassp}. Our results demonstrate that BAF-Net consistently outperforms single-modal baselines in terms of objective speech quality metrics.

\begin{figure*}
    \centering
    \includegraphics[width=\textwidth]{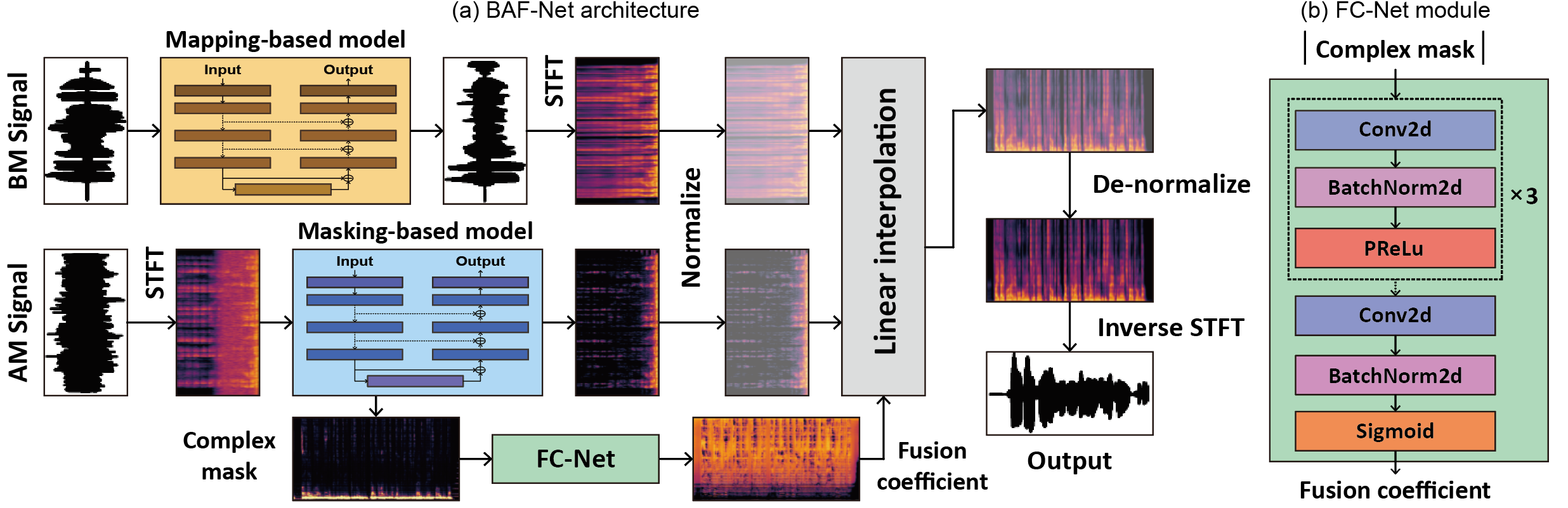}
    \caption{(a) Network structure of the proposed BAF-Net. (b) FC-Net module for the fusion coefficient estimation.}
    \label{fig:1}
\end{figure*}

\section{BAF-Net}

\subsection{Problem Settings}

The BAF-Net adaptively combines BMS and AMS using a noise-dependent fusion strategy. The fusion strategy is designed to select the enhanced BMS in regions dominated by severe noise, while leveraging denoised AMS in relatively cleaner regions. This allows the framework to exploit the inherent robustness of BMS against ambient noise and the rich spectral content provided by AMS in noise-free or low-noise conditions. The framework comprises two specialized networks:
\begin{enumerate}
    \item Mapping-based model ($f_{map})$ : Enhances raw BMS $x_{bm}$ by reconstructing high-frequency components lost due to tissue filtering, yielding the spectrogram $\hat{X}_{bm}\in \mathbb{C}^{T\times F}$.
    \begin{align}
        \hat{X}_{bm}=\text{STFT}(\hat{x}_{bm}),\quad \hat{x}_{bm}=f_{map}(x_{bm})
    \end{align}
    \item Masking-based model ($f_{mask})$ : Denoises AMS by estimating a complex ideal ratio mask (cIRM) $M\in \mathbb{C}^{T\times F}$. The denoised AMS spectrogram $\hat{X}_{am}$ is obtained by element-wise multiplication with the noisy AMS spectrogram $X_{am}$:
    \begin{align}
        \hat{X}_{am}=X_{am}\odot M,\quad M=f_{mask}(X_{am}).
    \end{align}
\end{enumerate}

\noindent The enhanced spectrograms from both modalities, $\hat{X}_{am}$ and $\hat{X}_{bm}$, are then combined using a linear interpolation scheme:
\begin{align}
    \hat{X}=\alpha\cdot \hat{X}_{am}+(1-\alpha)\cdot \hat{X}_{bm},
\end{align}
where the fusion coefficient $\alpha\in(0,1)^{T\times F}$ adaptively balances the enhanced BMS and the denoised AMS. The coefficient is estimated by the fusion coefficient network (FC-Net) $f_{F}$ using the cIRM magnitude $|M|$, which reflects the local SNR:
\begin{align}
    \alpha=f_{F}(|M|).
\end{align}
In strong noise regions, the FC-Net predicts $\alpha\rightarrow 0$, thereby favoring the enhanced BMS. Conversely, in lower noise regions, the FC-Net predicts $\alpha\rightarrow 1$, favoring the denoised AMS. This adaptive fusion approach ensures that the optimal modality is dynamically selected based on the instantaneous noise conditions at each TF bin.

To train the model, multi-resolution STFT \cite{yamamoto2020parallel} and L1 losses are jointly minimized:
\begin{align}
    \mathcal{L} = \mathbb{E}[\Vert \hat{x}-x\Vert_1] + \sum_{k=1}^K \mathbb{E}\left[\Vert\vert\text{STFT}_k(\hat{x})\vert-\vert\text{STFT}_k(x)\vert\Vert_1\right],
\end{align}
where $\text{STFT}_k$ computes spectral magnitudes with varying window/hop sizes. The L1 term minimizes waveform differences, while the multi-STFT loss matches spectral content across resolutions.

\begin{table*}[htbp]
  \centering
  \footnotesize
  \setlength{\tabcolsep}{2.2pt}
  \caption{Speech enhancement performance on the TAPS dataset at various SNR levels, evaluated via wideband PESQ, STOI, and CER.} 
  \label{tab:model_performance}
  \begin{tabular}{@{}ll*{15}{S[table-format=1.3]}@{}} 
    \toprule
    \multirow{2}{*}{Input} & \multirow{2}{*}{Model} & \multicolumn{5}{c}{PESQ} & \multicolumn{5}{c}{STOI} & \multicolumn{5}{c}{CER} \\
    \cmidrule(lr){3-7} \cmidrule(lr){8-12} \cmidrule(lr){13-17}
    & & {-20 dB} & {-10 dB} & {0 dB} & {10 dB} & {15 dB} & {-20 dB} & {-10 dB} & {0 dB} & {10 dB} & {15 dB} & {-20 dB} & {-10 dB} & {0 dB} & {10 dB} & {15 dB} \\
    \midrule
    BMS & SE-conformer \cite{kim2021se} & \multicolumn{2}{c}{---} & 1.971 & \multicolumn{2}{c}{---} & \multicolumn{2}{c}{---} & 0.892 & \multicolumn{2}{c}{---} & \multicolumn{2}{c}{---} & 0.244 & \multicolumn{2}{c}{---} \\
    \midrule
    \multirow{2}{*}{\makecell{Noisy\\AMS}} & SE-conformer \cite{kim2021se} & 1.116 & 1.305 & 1.706 & 2.109 & 2.240 & 0.523 & 0.715 & 0.839 & 0.893 & 0.900 & 0.886 & 0.736 & 0.465 & 0.308 & 0.275 \\
    & DCCRN \cite{hu2020dccrn}       & 1.114 & 1.259 & 1.610 & 2.062 & 2.289 & 0.498 & 0.656 & 0.768 & 0.811 & 0.822 & 0.861 & 0.699 & 0.440 & 0.261 & 0.243 \\
    \midrule
    \multirow{5}{*}{\makecell{Noisy\\AMS\\and\\BMS}} & FCN-LF \cite{yu2020time}      & 1.053 & 1.063 & 1.159 & 1.338 & 1.421 & 0.326 & 0.462 & 0.606 & 0.685 & 0.696 & 0.919 & 0.842 & 0.610 & 0.401 & 0.351 \\
    & FCN-EF \cite{yu2020time} & 1.228 & 1.345 & 1.413 & 1.464 & 1.482 & 0.775 & 0.809 & 0.828 & 0.842 & 0.845 & 0.578 & 0.483 & 0.430 & 0.390 & 0.381 \\
    & Attention-DC-CRN-LF \cite{wang2022attention}& 1.448 & 1.590 & 1.701 & 1.758 & 1.775 & 0.822 & 0.840 & 0.856 & 0.866 & 0.869 & 0.361 & 0.323 & 0.282 & 0.254 & 0.243 \\
    & Attention-DC-CRN-EF \cite{wang2022attention}& 1.546 & 1.789 & 2.000 & 2.127 & 2.156 & 0.840 & 0.871 & 0.891 & 0.902 & 0.905 & 0.386 & 0.290 & 0.225 & 0.197 & 0.185 \\
    & BAF-Net     & \textbf{2.082} & \textbf{2.260} & \textbf{2.491} & \textbf{2.751} & \textbf{2.875} & \textbf{0.904} & \textbf{0.914} & \textbf{0.929} & \textbf{0.944} & \textbf{0.949} & \textbf{0.222} & \textbf{0.206} & \textbf{0.187} & \textbf{0.170} & \textbf{0.167} \\
    \bottomrule
  \end{tabular}
\end{table*}

\subsection{BAF-Net architecture}
The architecture operates in the TF domain and consists of three core components: modality-specific processing networks for enhancing BMS and AMS, an FC-Net for noise-adaptive signal fusion, and a synthesis stage that reconstructs the final waveform. Fig. \ref{fig:1}a illustrates the BAF-Net architecture.

The BAF-Net initially processes BMS and AMS independently. An SE-conformer \cite{kim2021se} is used for BMS enhancement, while a DCCRN \cite{hu2020dccrn} is used for AMS denoising. The SE-conformer restores high-frequency information in BMS, and the DCCRN separates speech from noisy AMS, producing $\hat{X}_{bm}$ and $\hat{X}_{am}$, respectively. FC-Net estimates the fusion coefficient $\alpha$ using $|M|$.
Before fusion, $\hat{X}_{bm}$ and $\hat{X}_{am}$ are normalized using their respective mean energies $E(\hat{X}_{bm})$ and $E(\hat{X}_{am})$:
\begin{align}
    E(X) = \frac{1}{T \cdot F} \sum_{t=1}^{T} \sum_{f=1}^{F} \Bigl(\text{Re}\bigl(X(t,f)\bigr)^2 + \text{Im}\bigl(X(t,f)\bigr)^2\Bigr),
\end{align}
\begin{align}
    \tilde{X}_{bm}=\frac{\hat{X}_{bm}}{\sqrt{E(\hat{X}_{bm})}},\quad \tilde{X}_{am}=\frac{\hat{X}_{am}}{\sqrt{E(\hat{X}_{am})}}.
\end{align}
This normalization ensures that the fusion coefficient $\alpha$ balances the contributions of BMS and AMS based on their relative energy levels. The fusion coefficient $\alpha$ is then used to interpolate the normalized spectrograms linearly:

\begin{align}
    \tilde{X}=\alpha\cdot \tilde{X}_{am}+(1-\alpha)\cdot \tilde{X}_{bm}.
\end{align}
The fused output $\tilde{X}$ is then re-scaled using the average root energy of $\hat{X}_{bm}$ and $\hat{X}_{am}$ to restore the overall energy level:
\begin{align}
    \hat{X}=\tilde{X}\cdot\frac{\sqrt{E(\hat{X}_{bm})}+\sqrt{E(\hat{X}_{am})}}{2}.
\end{align}
Finally, the enhanced speech $\hat{x}$ is synthesized by applying inverse STFT to $\hat{X}$:
\begin{align}
    \hat{x}=\text{iSTFT}(\hat{X}).
\end{align}

For the mapping-based enhancement, we utilized the SE-conformer, an encoder-decoder architecture that combines convolutional neural networks with a Conformer-based sequence modeling network \cite{gulati2020conformer}. We adopted a variant of the SE-conformer \cite{kim2025tapsthroatacousticpaired}, optimized for BMS enhancement. For the masking-based denoising, we employed DCCRN, a complex-valued encoder-decoder network. DCCRN operates directly on complex spectrograms, enabling joint estimation of magnitude and phase through complex-valued operations. As illustrated in Fig.\ref{fig:1}b, the FC-Net adopts a convolutional architecture composed of three sequential blocks, each containing a Conv2D layer (16 channels, kernel size $7\times7$), Batch Normalization, and PReLU activation. The final layer replaces PReLU with a Sigmoid activation to constrain the fusion coefficient $\alpha$ to the range $[0,1]$.

\begin{figure*}
    \centering
    \includegraphics[width=\textwidth]{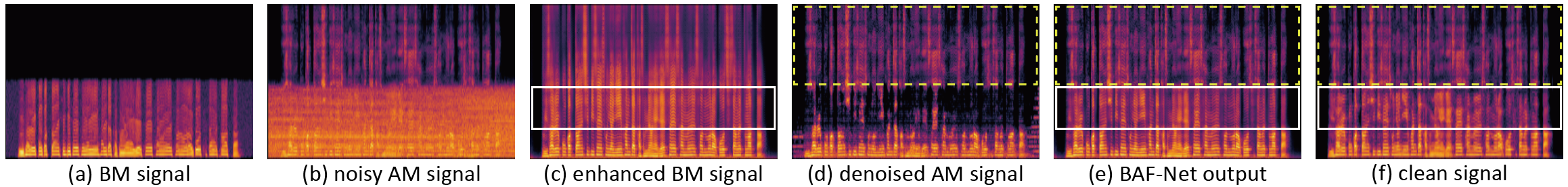}
    \caption{Spectrogram analysis of the enhancement process. (a) Body-conduction microphone signal, (b) noisy acoustic microphone signal, (c) enhanced body-conduction microphone signal, (d) denoised acoustic microphone signal, (e) final output, and (f) clean signal.}
    \label{fig:2}
\end{figure*}

\section{Experiments}
\subsection{Experiment on the TAPS dataset}
We evaluated the BAF-Net using the Throat-Acoustic Pairing Speech (TAPS) dataset \cite{kim2025tapsthroatacousticpaired}, which includes paired body-conduction microphone and acoustic microphone speech recordings. The dataset was divided into 4,000 training, 1,000 validation, and 1,000 evaluation pairs. To simulate noisy and reverberant conditions, we augmented the data with noise clips and room impulse responses (RIRs) from the DNS-2023 challenge dataset \cite{dubey2023icassp}. DNS-2023 provides 65,000 noise clips (150 classes) and 60,000 RIRs for simulating diverse acoustic environments. The noise clips and RIRs used in the test set were not included in the training set, ensuring that the evaluation accurately reflects the model's performance in unseen acoustic conditions.

For training, noisy AMS was synthesized by mixing clean AMS with a subset of DNS noise clips at SNRs ranging from $–15$ dB to 20 dB. To simulate reverberation, 75\% of training samples were convolved with a randomly selected subset of DNS RIRs, while the remaining 25\% retained anechoic conditions. Each training sample consisted of a triplet: BMS, noisy AMS, and clean AMS. For evaluation, we created a test set using 2,000 noise clips and 1,000 RIRs from DNS-2023 that were entirely excluded from training. Each evaluation sample was generated at five SNR levels: $–20$ dB, $–10$ dB, 0 dB, 10 dB, and 15 dB, with reverberation applied to 50\% of the samples. This resulted in 10,000 test utterances (2,000 noise clips × 5 SNRs, split evenly between reverberant and anechoic conditions).

\subsection{Training setup and baselines}
For comparison, several single-modal baseline models were trained:
\begin{itemize}
    \item \textbf{DCCRN} \cite{hu2020dccrn}: A masking-based model trained on noisy AMS.
    \item \textbf{SE-conformer} \cite{kim2021se}: A mapping-based model with two variants: one trained on noisy AMS and the other trained on BMS.
\end{itemize}
We also compared our BAF-Net with the following multi-modal baselines in previous studies:
\begin{itemize}
    \item \textbf{FCN~\cite{yu2020time}}: A time-domain Fully Convolutional Network (FCN) utilizing two fusion strategies:
    \begin{itemize}
        \item \textit{Early Fusion (EF)}: Concatenates BMS and AMS at the input stage and processes the concatenated signal with the FCN.
        \item \textit{Late Fusion (LF)}: Processes BMS and AMS independently through separate FCNs and subsequently processes the concatenated output with another FCN.
    \end{itemize}
    \item \textbf{Attention-DC-CRN~\cite{wang2022attention}}: A complex-domain Densely Connected Convolutional Recurrent Network (DC-CRN) employing an attention mechanism in two fusion strategies:
    \begin{itemize}
        \item \textit{Early Fusion (EF)}: Integrates AMS and BMS spectrograms at the input stage using attention mechanisms, followed by the DC-CRN module.
        \item \textit{Late Fusion (LF)}: Processes each modality through separate DC-CRN modules and combines the outputs with an attention-based fusion.
    \end{itemize}
\end{itemize}
All the models, with the exception of Attention-DC-CRN, were trained using a 25 ms window length, 6.25 ms hop size, and 512-point FFT for spectrogram extraction. Attention-DC-CRN was trained using a 32 ms window length, 16 ms hop size, and 512-point FFT for spectrogram extraction. The Adam optimizer was used with a learning rate of $3 \times 10^{-4}$ and momentum parameters $\beta_{1}=0.9$ and $\beta_{2}=0.99$. 

For the BAF-Net, the mapping and masking models were pre-trained separately for 200 epochs. The pre-trained weights were then transferred into the BAF-Net, and the full model was fine-tuned for another 200 epochs. All models, including baselines, employed a combined multi-resolution STFT loss ($K=3$) and L1 loss, with multi-STFT configurations set to $\{512,50,240\}$, $\{1024,120,600\}$, and $\{2048,240,1200\}$ (FFT size, window size, and frame shift). Early stopping based on the validation set ensured a fair comparison across architectures. The model architecture and training code for BAF-Net, as well as baseline models, are available at the public repository\footnote{https://github.com/yskim3271/BAF-Net}.

\section{Experimental results and discussion}
The BAF-Net was evaluated using the simulated TAPS dataset and three metrics. First, we used the wideband Perceptual Evaluation of Speech Quality (PESQ) \cite{rix2001perceptual}, which ranges from $-0.5$ to $4.5$. Second, we measured the Short-Time Objective Intelligibility (STOI) \cite{taal2010short}, which ranges from 0 to 1. Finally, to quantify the recovery of speech information, we assessed the Character Error Rate (CER) using the Whisper-large-v3-turbo automatic speech recognition (ASR) model \cite{radford2022whisper}, which had been fine-tuned\footnote{https://huggingface.co/ghost613/whisper-large-v3-turbo-korean} on the Korean Zeroth dataset\footnote{https://openslr.org/40}. Enhanced speech samples were transcribed using the ASR model, and CER was computed by comparing transcriptions with ground-truth labels.

As shown in Table 1, at low SNRs ($–20$ dB to 0 dB), single-modal BMS-based models outperformed AMS-based models, as noise suppression in AMS inevitably removes speech components. Conversely, AMS-based models achieved superior results at high SNRs (10 and 15 dB), likely due to the spectral limitations of BMS. The BAF-Net outperformed single-modal models across all noise levels, demonstrating the effectiveness of its adaptive fusion in leveraging the complementary strengths of both modalities. In addition, the BAF-Net exhibited clear advantages compared to other multi-modal baselines. At extremely low SNR conditions ($-20$ and $-10$ dB), other multi-modal approaches showed limited improvements due to severe noise contamination in AMS. In contrast, BAF-Net leveraged BMS’s inherent noise resistance, resulting in significant performance gains. At higher SNRs (10 and 15 dB), the adaptive fusion mechanism utilized AMS’s spectral richness, leading to notable enhancements in performance.

Spectrogram analysis of the enhanced speech demonstrates BAF-Net's ability to effectively utilize modality-specific strengths. Fig. 2a displays the spectrogram of the BMS input. For a noisy AMS containing white noise in the low frequency region (Fig. 2b), the mapping-based model restored high-frequency components in BMS (Fig. 2c). In contrast, the masking-based model effectively suppressed noise in AMS (Fig. 2d). The fused output (Fig. 2e) seamlessly integrates enhanced BMS and denoised AMS through adaptive fusion. In the noise-free region (Fig. 2, yellow dashed boxes), the denoised AMS retained nearly noise-free characteristics, causing the fusion mechanism to prioritize AMS. As a result, the fusion mechanism produced an output close to the clean reference. Conversely, the fusion mechanism favored the enhanced BMS in the noise-dominant region (Fig. 2, white boxes).

We analyzed the cIRM magnitude and fusion coefficient at 10 dB and $–20$ dB to verify the adaptive fusion. The noisy AMS spectrograms are shown in Fig. 3a and Fig. 3b. At 10 dB, the mask selectively attenuated noise-dominant regions (Fig. 3c), while the fusion coefficient prioritized AMS in clean areas and BMS in noisy areas (Fig. 3e). At $–20$ dB, the mask suppressed noise across broader regions (Fig. 3d), and the fusion coefficient depended primarily on BMS except in high-frequency bins (Fig. 3f). This dynamic fusion confirms the BAF-Net’s ability to compensate for information loss in noisy regions using BMS while preserving AMS fidelity in cleaner areas.

\begin{figure}[t]
    \centering
    \includegraphics[width=\linewidth]{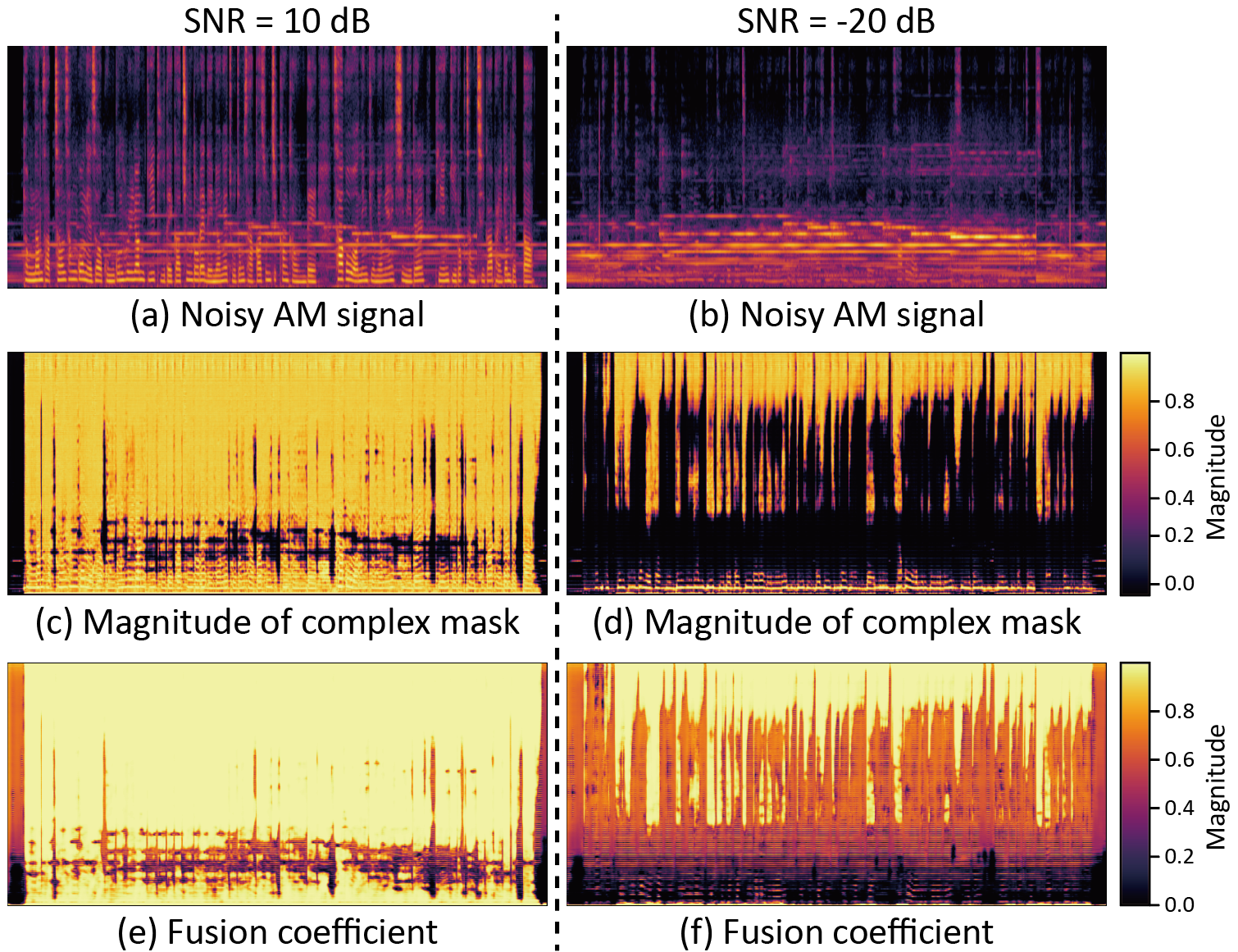}
    \caption{Visualization of the fusion mechanism in the BAF-Net. (a, b) Noisy acoustic microphone signal inputs, (c, d) complex mask magnitudes estimated by the masking-based model, and (e, f) fusion coefficients derived from the complex mask magnitudes at SNR levels of 10 dB and $-20$ dB, respectively.}
    \label{fig:3}
\end{figure}

\section{Conclusion}

In conclusion, this study presents the BAF-Net, a multi-modal framework that combines BMS and AMS using task-specific enhancement and adaptive fusion. It restores high-frequency components in BMS using a mapping-based network and reduces noise in AMS using a masking-based model. This approach overcomes the weaknesses of each modality while maximizing their strengths. The BAF-Net employs a simple yet effective fusion method, utilizing enhanced BMS for noisy segments and denoised AMS for clearer parts. Experimental results across various SNR levels show that the BAF-Net surpasses previous single- and multi-modal models, highlighting the importance of modality-specific processing and adaptive fusion.

\section{Acknowledgements}

This research was supported by the National Research Foundation of Korea grants (RS-2025-00516311, RS-2020-NRF047144) and by the Institute of Information \& Communications Technology Planning \& Evaluation grant (RS-2019-II191906, Artificial Intelligence Graduate School Program).

\bibliographystyle{IEEEtran}
\bibliography{mybib}

\end{document}